\documentclass[twoside,twocolumn]{article}

\usepackage{blindtext} %

\usepackage[sc]{mathpazo} %
\usepackage[T1]{fontenc} %
\usepackage{microtype} %

\usepackage[english]{babel} %

\usepackage[hmarginratio=1:1,top=32mm,columnsep=20pt]{geometry} %
\usepackage[hang, small,labelfont=bf,up,textfont=it,up]{caption} %
\usepackage{booktabs} %

\usepackage{lettrine} %

\usepackage{enumitem} %
\setlist[itemize]{noitemsep} %

\usepackage{abstract} %

\usepackage{titlesec} %
\renewcommand\thesection{\Roman{section}} %
\renewcommand\thesubsection{\roman{subsection}} %
\titleformat{\section}[block]{\large\scshape\centering}{\thesection.}{1em}{} %
\titleformat{\subsection}[block]{\large}{\thesubsection.}{1em}{} %

\usepackage{fancyhdr} %
\usepackage{titling} %

\usepackage{hyperref} %
\usepackage{import}
\usepackage{bm}
\usepackage{optidef}
\usepackage{units}
\usepackage{times} %
\usepackage{amsmath} %
\usepackage{amssymb}  %
\usepackage{graphicx} %
\usepackage{units}
\usepackage{multirow}
\usepackage{booktabs}
\usepackage{colortbl}
\usepackage{makecell}
\usepackage{algorithmicx}
\usepackage{algpseudocode}
\usepackage{pifont}
\usepackage{hhline}
\usepackage[export]{adjustbox}
\usepackage[sort&compress,numbers]{natbib}        %
\usepackage{stfloats}		%
\usepackage{accents}
\usepackage{color,soul}
\usepackage{xcolor}
\usepackage{lipsum}

\RequirePackage{pgfplots}
\pgfplotsset{compat=1.12}
\graphicspath{{images/}}	%
\usepackage{svg}
\usepackage{pgfplots}

\definecolor{mygreen}{RGB}{0, 176, 73}

\pretitle{\begin{center}\Huge\bfseries} %
\posttitle{\end{center}} %
\title{Learning-Based Model Predictive Control for the Energy Management of Hybrid Electric Vehicles Including Driving Mode Decisions} %
\author{%
\textsc{David Theodor Machacek, Stijn van Dooren, Thomas Huber, Christopher Onder}\thanks{D. T. Machacek, S. v. Dooren, C. H. Onder are with the Institute for Dynamic Systems and Control (IDSC), ETH Z\"urich, ZH, 8092, Switzerland (e-mail: {davidm}@ethz.ch)} \thanks{T. Huber is with the Powertrain Solutions and System Engineering Group, Robert Bosch GmbH, 70442 Stuttgard, Germany}\\[1ex] %
\normalsize ETH Z\"urich, Switzerland \\ %
\normalsize \href{mailto:davidm@ethz.ch}{davidm@ethz.ch} %
}
\date{\today} %
\renewcommand{\maketitlehookd}{%
\begin{abstract}
This paper presents an online-capable controller for the energy management system of a parallel hybrid electric vehicle based on model predictive control. Its task is to minimize the vehicle's fuel consumption along a predicted driving mission by calculating the distribution of the driver's power request between the electrical and the combustive part of the powertrain, and by choosing the driving mode, which depends on the vehicle's clutch state. %
The inclusion of the clutch state in a model predictive control structure is not trivial because the underlying optimization problem becomes a mixed-integer program as a consequence. Using Pontryagin's Minimum Principle and a simplified vehicle model, it is possible to prove that a drive cycle-dependent critical power request $P_\mathrm{crit}$ exists, which uniquely separates the optimal driving mode. Based on this result, a learning algorithm is proposed to determine $P_\mathrm{crit}$ during the operation of the vehicle. The learning algorithm is incorporated into a multi-level controller structure and 
the working principle of the resulting multi-level learning-based model predictive controller is analyzed in detail using two realistic driving missions. A comparison to the solution obtained by Dynamic Programming reveals that the proposed controller achieves close-to-optimal performance. 
\end{abstract}
}

\begin{document}

\maketitle

\section{Introduction} \label{sec:Introduction}
\begin{figure}[h!]
	\centering
	\includegraphics[scale=0.6]{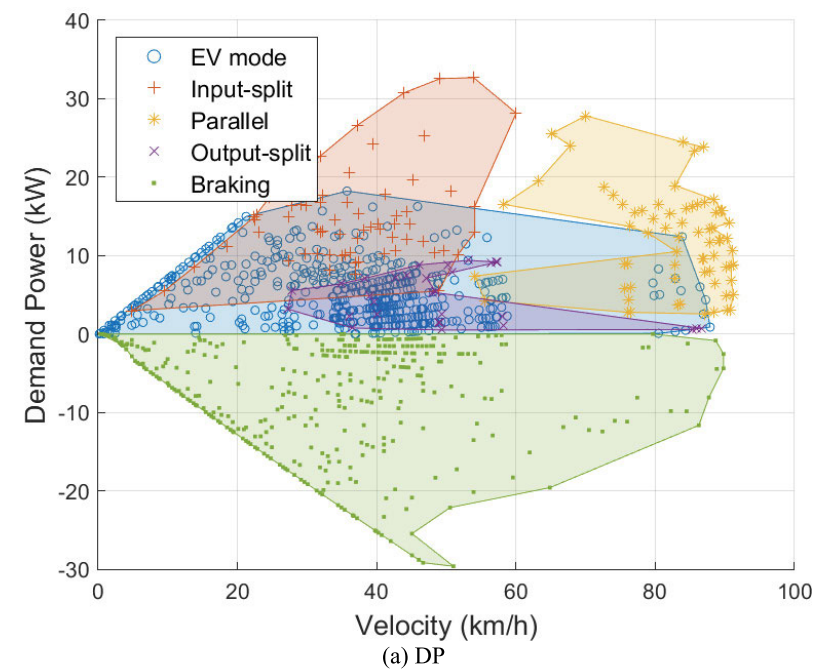}
	\caption{Optimal distribution of the driving modes for the multimode HEV that was investigated and published by the author of \cite{kwon2022control}. The optimal driving modes are calculated using the DP algorithm and are displayed in the ($P_\mathrm{req}, v$)-space.}
	\label{fig:controlMap_plane}
\end{figure}

Hybrid electric vehicles (HEV) are a common way to cope with the ever more stringent CO$_2$ emission regulations. In addition to an internal combustion engine (ICE), HEVs include one or more electric motors (EM) connected to an electric storage, typically a battery. On the one hand, this second energy source offers an additional degree of freedom for power distribution \cite{guzzella2013vehicle}. On the other hand, for the case of parallel-hybrids, the clutch can be used to distinguish between two driving modes; namely HEV mode if the clutch is closed and the engine is on, and EV mode if the clutch is open and the engine is off. In EV mode, the engine is decoupled from the rest of the drivetrain to avoid its substantial pumping and friction losses.\\
\\
The optimal choice of the driving mode for a multimode HEV was investigated by the authors of \cite{kwon2022control}, and they presented a typical distribution of the optimal driving mode, in a space spanned by the driver's power request and the vehicle speed (($P_\mathrm{req}, v$)-space), as shown in Figure \ref{fig:controlMap_plane}. The distribution of a driver's power request between the individual propulsion components is usually referred to as the task of an energy management system (EMS). The inclusion of the degree of freedom of choosing the driving mode as part of the EMS controller is discussed in this text.
\\
\\
The task to find the best possible control inputs that optimize a control-goal, while respecting a wide variety of constraints, can be formulated as an optimal control problem (OCP). The inclusion of continuous optimization variables (such as component powers), as well as discrete optimization variables (such as the driving mode) leads, in the most general case, to the formulation of a nonlinear mixed-integer OCP. This group of OCPs is notoriously hard to solve and usually requires a lot of computation time.\\
\\
The following literature review is two-fold. First, the most recent approaches to solve the mixed-integer OCP that results from the incorporation of the discrete variables in the standard EMS are summarized in order to frame the novel control concept of this publication. Second, the method of MPC is highlighted as a powerful controller concept with an emphasis on the multi-level control structure to enable the online-capability of MPC for highly dynamic control systems.
\subsection{EMS with Driving Modes}
To investigate potential fuel savings and to assess the performance of online-capable EMS controllers, so called optimal solutions are calculated. The Dynamic Programming (DP) algorithm \cite{bellman1966dynamic} is commonly used, as it is able to cope with a wide variety of problem formulations while providing the optimal solution \cite{wang2012dynamic}. The drawback of DP is the curse of dimensionality \cite{powell2007approximate}, which can prohibit the use of its applicability to large OCPs. To tackle this problem, the authors of \cite{robuschi2019minimum} presented a new approach to solve the mixed-integer OCP, including the engine ON/OFF mode, based on iterative linear programming. In their most recent publication \cite{robuschi2021multiphase}, they present an algorithm to approximate the multiphase mixed-integer nonlinear OCP by solving a sequence of nonlinear programs and mixed-integer linear programs.\\
\\
Online-capable control methods for the EMS have to calculate the control parameters much faster than offline benchmark methods. This is usually achieved by using extensive data driven methods, or with the help of pre-computed control maps to decide the driving mode. Both methods are highlighted in the following.\\
\\
Chen et. al. \cite{chen2022neural} employ Pontryagin's Minimum Principle (PMP) \cite{pontryagin1962grv} to solve the EMS including driving modes and suggest the use of two neural networks trained on optimal results calculated offline using DP. A Bayesian regularized neural network is trained to estimate the optimal battery costate, and a backpropagation neural network is trained to estimate the optimal engine ON/OFF mode. During online control, both networks are used to estimate the battery costate and correct it using a heuristic based on the prediction of the engine ON/OFF mode. Subsequently, the corrected battery costate is passed on to a controller that is based on the well-known equivalent consumption minimization strategy (ECMS) \cite{paganelli2000simulation}.\\
A different approach to include the engine ON/OFF mode in the EMS is presented in \cite{guo2018energy}. The authors propose the use of an evolutionary algorithm to calculate the optimal battery state of charge (SoC) reference trajectory for a plug-in HEV including the engine ON/OFF mode. The theory of PMP and model simplifications were leveraged to find an analytic expression for the engine ON/OFF command and to simultaneously reduce the number of optimization parameters of the evolutionary algorithm. With the pre-optimized SoC trajectory and an adaptive ECMS controller the authors show good performance during a hardware-in-the-loop experiment.\\
\\
The authors of the following publications present methods to compute and/or use offline generated control maps for online control. The author of \cite{kwon2022control} uses DP to perform offline optimizations of the mixed-integer OCP on different drive cycles to find the optimal distribution of the driving mode for a Dual Split Hybrid vehicle. Subsequently, different machine learning techniques were trained on the results obtained by DP to identify correlations between the vehicle speed, the power request, the torque request, and the driving mode. It was shown that the best separation of the optimal driving modes could be achieved in the ($P_\mathrm{req}, v$)-space, which was then stored in a control map.\\
A similar approach is used by the authors of \cite{mashadi2020fuel} for a serial-parallel HEV. They use DP optimizations and subsequently a particle swarm optimization to create a control map for the driving modes in the ($P_\mathrm{req}, v$)-space. An additional heuristic is applied that incorporates the SoC to ensure that the battery state constraints are satisfied. The method was extended in \cite{khademnahvi2022introducing} to also include predictive drive mission data.
\subsection{Model Predictive Control for the EMS}
Model predictive control is a powerful control technique that has multiple advantages for online control of the EMS of HEVs. On the one hand, it allows for the inherent incorporation of input constraints, and state constraints, as well as the possibility to incorporate predictive data for optimal trajectory planning. On the other hand, multi-objective control, i.e., the pursuit of multiple conflicting control objectives, is straightforward and doesn't require offline pre-tuned weights to describe their trade-off. The drawback of MPC is its often limiting computation time, because an optimization problem needs to be solved to calculate the controller inputs. The authors of \cite{van2019multi} state that an EMS controller should be able to calculate the power-split with an update frequency of roughly 10 Hz. Therefore, a mixed-integer OCP usually prohibits the use of MPC for online control. However, the solution to such problems can be approximated more efficiently by splitting the mixed-integer OCP into two optimization subproblems of lower computational burden. 
A typical approach would be to solve the subproblems using an iterative scheme (\cite{nuesch2014convex}, \cite{robuschi2019minimum}), or to organize the subproblems in multiple control layers.
In multi-level control, a higher-level controller usually acts as a supervisory controller with a slower execution time, and the lower-level controller computes the control inputs in real-time. Some of the recent publications on the topic of multi-level MPC for the EMS of HEVs are presented in the following.\\
\\
In \cite{uebel2017optimal} and \cite{uebel2019two}, Uebel et. al. present a two-level MPC approach that leverages Sequential Quadratic Programing and a combination of DP and PMP to solve the EMS for an adaptive cruise control scenario. The higher-level controller is used to calculate setpoints for the lower-level controller. The engine ON/OFF mode, as well as the gear choice, are computed in the lower-level controller using the setpoints.\\
\\
The authors of \cite{razi2021predictive} also present a two-level MPC control approach for the EMS of an HEV that requires solving a mixed-integer OCP. The multi-level structure is based on a DP-PMP approach in the supervisory control layer and a linear quadratic tracking method in the lower level controller.\\
\\
Aubeck et. al. \cite{aubeck2022generic} present a two-level MPC scheme that leverages the DP algorithm and a generic stochastic particle filter-based algorithm to include the engine ON/OFF mode in the EMS. The supervisory controller is a DP algorithm that provides target set-point values to the lower-level particle filter-based algorithm, which is also executed using an MPC scheme and calculates the power split and the demanded engine ON/OFF mode, in real-time.%
\subsection{Contributions}
In several recent publications (\cite{kwon2022control} \cite{mashadi2020fuel} \cite{khademnahvi2022introducing}), a control map, which is precalculated offline and defined in the \mbox{($P_\mathrm{req},\, v$)-space}, is proposed to choose the driving mode. In contrast, however, in this paper, a novel approach is presented that uses a similar control map as part of a learning-based MPC (LB-MPC) algorithm to solve the EMS including the driving modes for a parallel HEV.\\
The control map:
\begin{itemize}
	\item Is defined in the ($P_\mathrm{req},\, v$)-space, and used to estimate the future optimal driving mode for an MPC. The actual driving mode is calculated based on optimal control theory.
	\item Is shown to depend on the driving mission. Therefore, a learning algorithm (LA) is used to learn the optimal control map, using the driving mode choice during online operation of the vehicle.
\end{itemize}
Similar to \cite{murgovski2013engine}, the mixed-integer OCP is approximated by splitting it, using convex optimization and PMP. However, instead of using an iterative approach, the two subproblems are arranged in a multi-level controller structure. On the higher-level, a convex optimal control problem  is formulated and solved via an MPC scheme. On the lower-level, a Hamiltonian function is formulated, based on the costates of the MPC. It calculates the optimal power split, and the optimal driving mode in real-time.
\\
\subsection{Structure}
This paper is organized as follows: In \mbox{Section \ref{sec:Problem Description}}, the HEV drivetrain is outlined and the mixed-integer OCP is formulated. Subsequently, a simplified drivetrain model is shown as well as a convex reformulation of the mixed-integer OCP is presented. Finally, the investigated drive cycles are presented. In \mbox{Section \ref{sec:driving_mode_analysis}}, the optimal driving mode distribution is discussed based on the theory of PMP for a fixed vehicle operating point, and the findings are extended to the full vehicle model. In \mbox{Section \ref{sec:controller_structure}}, the LB-MPC controller structure is presented in detail. In \mbox{Section \ref{sec:learning_algorithm}}, the LA is presented and its working principle is demonstrated. In \mbox{Section \ref{sec:online_implementation}}, practical issues for the online implementation of the LB-MPC are highlighted, and robustifying methods are discussed to ensure the feasibility during online control. In \mbox{Section \ref{sec:case_study}}, the LB-MPC's performance is assessed in two test scenarios. In \mbox{Section \ref{sec:outlook}}, an outlook on future research is presented.
\section{Problem Description} \label{sec:Problem Description}
In this section, the HEV powertrain is introduced and the optimal control problem is outlined. A more detailed powertrain description is presented in our previous work \cite{machacek2022multi}.
\subsection{System Description} \label{subsec:System Description}
\begin{figure}[h!]
\centering
   \def\svgwidth{1\columnwidth}
{\normalsize
	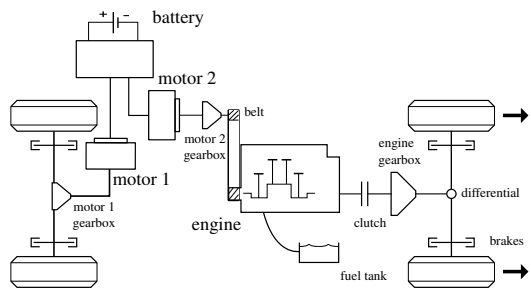}
\caption{Schematic illustration of the P$_4$/P$_0$ parallel hybrid drivetrain.}
\label{fig:power_train}
\end{figure}

Figure \ref{fig:power_train} shows a physical schematic of the powertrain architecture of the vehicle considered in this work. It features a P4 parallel-hybrid powertrain with an electric motor, which is hereafter referred to as motor 1, that is connected to the rear axle. The combustion engine is connected to the front axle through a seven-speed gearbox and the clutch. A belt driven P0 starter/generator electric motor, which is referred to as motor 2, is permanently connected to the engine crankshaft. Both electric motors feature a fixed gear transmission ratio and can draw from or feed power to the battery. The vehicle parameters are listed in Table \ref{tab:veh_pars}.
\begin{table}[h!]
\vspace{-0.2cm}
\centering
\caption{Vehicle parameters of the P4/P0 parallel HEV.}
\label{tab:veh_pars}
\begin{tabular}{l| l| l}\hhline{-|-|-}
\multirow{2}{*}{Vehicle} & Type & Premium sedan \\
 & Mass & $\unit[1917]{kg}$\\\hhline{-|-|-}
\multirow{2}{*}{Engine} & Type & 4-cyl. gasoline  \\
 & Max. power & $\unit[146]{kW}$ \\\hline 
\multirow{2}{*}{Motor 1} & Type & PMSM  \\
 & Rated power & $\unit[20]{kW}$ @4500\unit{rpm} \\\hline
\multirow{2}{*}{Motor 2} & Type & Claw pole machine  \\
 & Rated power & $\unit[10]{kW}$ @4000\unit{rpm} \\\hline
\multirow{2}{*}{Battery} & Battery voltage & \unit[48]{V} \\
 & Battery capacity & \unit[770]{Wh} \\\hline
\end{tabular}
\end{table}\\
The driver's power request $P_\mathrm{req}$ is divided between the rear axle and the front axle using a power split device, which is denoted with $u_1$. The power request for the front axle is divided between the combustion engine and motor 2 using a second power split device, which is denoted with $u_2$. All power flows are explained in the following.
\\
\\
The motor power and battery power subscripts underlie the following notation:
\begin{equation} \label{eq:source_powers}
	\begin{aligned}
		P_{s_{i}} &= P_{m_{i}} + P_{l_{i}} \quad i \in \{1,2\} \\
		P_{s_b} &= P_b + P_{l_b}.
	\end{aligned}
\end{equation}
The motor source power $P_{s_i}$ is a summation of the mechanical motor power $P_{m_i}$ and the motor loss power $P_{l_i}$. The battery source power $P_{s_b}$ is a summation of the battery power $P_{b}$ and the battery loss power $P_{l_b}$.
\subsection{Powertrain Model}\label{subsec:powertrain_model}
The modeling approach used in this work is based on \cite{guzzella2013vehicle} and is outlined in the following.\\
\\
\textit{Propulsion system models:}\\
The engine's fuel consumption is obtained using a map that was evaluated on a test bench and depends on the engine speed $\omega_{e}$ and the engine torque $T_{e}$:
\begin{equation} \label{eq:fuel_map}
\dot{m}_f = f(\omega_{e}, T_e).
\end{equation}
Considerable engine power losses $P_{l_e}$ occur when the engine is motored. Based on \cite{guzzella2009introduction}, a high-fidelity speed-dependent component model is used to combine friction losses and gas-exchange losses:
\begin{equation}
	P_{l_e} = f(\omega_{e}).
\end{equation}
The electric motor losses are stored in maps that are derived from high-fidelity component models:
\begin{equation}
	P_{l_i} = f(T_{m_i}, \omega_{i}, V_{\mathrm{oc}}).
\end{equation}
They depend on the motor torque $T_{m_i}$, the motor speed $\omega_{i}$, and the battery open circuit voltage $V_{\mathrm{oc}}$.\\
\\
The battery is modeled using a Thévenin equivalent model \mbox{\cite{guzzella2013vehicle}}. The current $I_b$ and battery internal losses $P_{l_b}$ that result from a requested battery power $P_b$ are calculated as
\begin{equation} \label{eq:thevenin}
\begin{aligned}
P_b &= P_{s_1}+P_{s_2}+P_{\mathrm{aux}} \\
I_b &= \frac{V_{oc}-\sqrt{V_{oc}^2-4R_i P_b}}{2R_i} \\
P_{l_b} &= R_i \cdot I_b^2,
\end{aligned}
\end{equation}
where the inner resistance $R_i$ and the auxiliary losses $P_{\mathrm{aux}}$ are assumed to be constant. The battery open circuit voltage $V_{\mathrm{oc}}$ can be approximated to depend quadratically on the state of charge \cite{zhang2016generalized}.\\
\\
The battery's internal energy is calculated as:
\begin{equation} \label{eq:battery_Eb}
	E_b = \mathrm{SoC} \cdot Q_\mathrm{max}\cdot V_\mathrm{oc},
\end{equation}
where $Q_\mathrm{max}$ is the full battery capacity.
Together with (\ref{eq:source_powers}), the battery energy dynamics are described by
\begin{equation} \label{eq:soc_dynamics}
\begin{aligned}
	\dot{E_b} &= -P_{s_b}(P_b, V_\mathrm{oc}).
\end{aligned}
\end{equation}
\textit{Power split:}\\
The hybrid electric powertrain can be operated in three distinct driving modes, which are defined by the combined \mbox{engine-clutch-input} $u_3 \in \{1,2,3\}$, and are listed in \mbox{Table \ref{tab:driving_modes}}. If the clutch is closed and the engine is running, the vehicle is in HEV mode. If the clutch is open and the engine is off, the vehicle is in EV mode. The powertrain can also be operated in a third mode, which is called Recuperation mode, with closed clutch and the engine being motored. This results in engine friction losses, but gives access to \mbox{motor 2} for additional regenerative braking power. 
Also, the clutch can be open and the engine ON, which allows to recharge the battery during standstill (e.g. in traffic). However, the use of this combination is trivial and also not relevant for the control problem at hand, i.e., to distribute the driver's power request optimally.
The combined engine-clutch-input can be separated into binary variables for the clutch $c_0\in \{0,1\}$ and the engine $e_0\in \{0,1\}$. The binary variable $c_0=1$ stands for clutch engaged, and $c_0=0$ stands for clutch open. The binary variable $e_0=1$ stands for engine ON, and $e_0=0$ stands for engine OFF.
\begin{table}[h!]
\vspace{-0.2cm}
\centering
\caption{Driving modes of the P4/P0 parallel HEV.}
\label{tab:driving_modes}
\begin{tabular}{l| l| l}\hhline{-|-|-}
HEV mode & $c_0=1$, $e_0=1$ & $u_3=1$\\
Recuperation mode & $c_0=1$, $e_0=0$ & $u_3=2$\\
EV mode & $c_0=0$, $e_0=0$ & $u_3=3$\\\hline
\end{tabular}
\end{table}\\
The requested force at the wheels is a function of the combined clutch and engine state $u_3$, the aerodynamic drag force and rolling resistance $F_d$, the gravitational force $F_g$, and the inertial force to accelerate the vehicle $F_i$:
\begin{equation} \label{eq:vehiclePower}
	\begin{aligned}
		&F_{\mathrm{req}} =\\
		&=\begin{cases}
			F_d(v) + F_i(a,\omega_{w}, \omega_{1}, \omega_{2}, \omega_{e}) + F_g(\Gamma)& \text{$u_3 = 1$}\\
			F_d(v) + F_i(a,\omega_{w}, \omega_{1}, \omega_{2}, \omega_{e}) + F_g(\Gamma) + \nicefrac{P_{l_e}}{\omega_{e}}& \text{$u_3 = 2$}\\
			F_d(v) + F_i(a,\omega_{w}, \omega_{1}) + F_g(\Gamma)& \text{$u_3 = 3$},\\
		\end{cases}
	\end{aligned}
\end{equation}
where $\Gamma$ denotes the road gradient, $v$ the velocity, and $a$ the vehicle acceleration. If the engine is motored, the engine power losses are added to the power request.\\
The resistive force $F_d$ is approximated using a quadratic dependency on $v$. The inertial force $F_i$ is calculated by accelerating the total mass of the vehicle $m_\mathrm{tot}$, which is a function of the vehicle's mass $m$ and an equivalent mass of all rotating parts $m_\mathrm{rot}$ \cite{guzzella2013vehicle}:
\begin{equation} \label{eq:mtot}
	\begin{aligned}
		m_\mathrm{tot} &= m + m_\mathrm{rot}(\omega_w, \omega_1, \omega_2, \omega_e)\\
		F_i &= m_\mathrm{tot} \cdot a,
	\end{aligned}
\end{equation}
where $\omega_w$, $\omega_1$, $\omega_2$, $\omega_e$ are the speeds of the wheels, \mbox{motor 1}, \mbox{motor 2}, and the engine, respectively. When the clutch is open, the contributions of $\omega_e$ and $\omega_2$ are zero.\\
For simplicity of the following equations, a reformulation to the power domain is made:
\begin{equation} \label{eq:force_to_power}
P_{\mathrm{req}} = F_{\mathrm{req}} \cdot v.
\end{equation}
The first power split defines the mechanical power demand for motor 1:
\begin{equation} \label{eq:u1}
P_{m_1} = P_{\mathrm{req}} \cdot u_1.
\end{equation}
The second power split defines the mechanical power demand for motor 2. The constant engine gearbox efficiency $\eta_{\mathrm{GB}}$ leads to the following power flow directional dependency:
\begin{equation} \label{eq:u2}
P_{m_2} = 
	\begin{cases}
		\nicefrac{P_{\mathrm{req}}}{\eta_{\mathrm{GB}}} \cdot (1-u_1) \cdot u_2 & \text{$P_{\mathrm{req}}\cdot (1-u_1) \geq 0$}\\
		P_{\mathrm{req}} \cdot \eta_{\mathrm{GB}} \cdot (1-u_1) \cdot u_2 & \text{$P_{\mathrm{req}}\cdot (1-u_1) < 0$}.
	\end{cases}
\end{equation}
In addition to defining the power of motor 2, $u_2$ is also used to define the mechanical engine power $P_{m_e}$ and the friction brake power $P_{m_\mathrm{brk}}$
\begin{equation} \label{eq:Pbrk}
\begin{aligned}
P_{m_e} & = 
\begin{cases}
\frac{P_{\mathrm{req}}\cdot (1-u_1)\cdot(1-u_2)}{\eta_{\mathrm{GB}}} \, \hspace{1.4cm} \text{$P_{\mathrm{req}}\cdot (1-u_1) \geq 0$} \\
0 \, \hspace{3.75cm} \text{$P_{\mathrm{req}}\cdot (1-u_1) < 0,$}
\end{cases} \\
P_{m_\mathrm{brk}} & = 
\begin{cases}
0 \, \hspace{3.75cm} \text{$P_{\mathrm{req}}\cdot (1-u_1) \geq 0$} \\
P_{\mathrm{req}} \cdot (1-u_1)\cdot(1-u_2) \, \hspace{0.3cm} \text{$P_{\mathrm{req}}\cdot (1-u_1) < 0$}.
\end{cases}
\end{aligned}
\end{equation}
To facilitate the formulation of a power balance in the optimal control problems, the description of the engine power and the motor 2 power at the level of power split $u_1$ are defined as
\begin{equation} \label{eq:powers_gearbox1}
	\begin{aligned}
		P_{\mathrm{GB}_e} &= P_{m_e} \cdot \eta_{\mathrm{GB}}\\
		P_{\mathrm{GB}_2} &= 
		\begin{cases}
			P_{m_2} \cdot \eta_{\mathrm{GB}} \quad \quad \quad &P_{m_2} \geq 0 \\
			\frac{P_{m_2}}{\eta_{\mathrm{GB}}} \quad \quad \quad &P_{m_2} < 0.
		\end{cases} 
	\end{aligned}
\end{equation}
Motor 2 can boost as well as be used for regenerative braking, resulting in a dependence on the power flow.\\
\\
\textit{Optimal Control Problem:}\\
Defining the input vector $\mathbf{u} \coloneqq [u_1, u_2, u_3]^T$, the optimal control problem of the EMS including optimal driving mode selection can be formulated. Note that, for the sake of readability (\ref{eq:vehiclePower})-(\ref{eq:powers_gearbox1}) are omitted in the following:
\begin{subequations}\label{eq:OCP}
\begin{alignat}{2}
\underset{\mathbf{u}}{\text{min}} \quad \quad \, & \int_{t_0}^{t_{\text{f}}}  \dot{m}_f \, dt\\
\text{   \qquad s.t.} \quad \dot{E}_b & = -P_{s_b}(P_b, V_\mathrm{oc}) \label{eq:OCP_state}\\
E_b(t_0) & = E_{b_0}\\
E_b(t_f) & \geq E_b(t_0)\\
P_{\text{GB}_2} & = P_{\text{req}}-P_{m_1} - P_{\text{GB}_e} - P_{\text{GB}_\text{brk}}\\
P_{1_\text{min}} & \leq P_{m_1} \leq P_{1_\text{max}}\\
P_{2_\text{min}}\cdot c_0 & \leq P_{m_2} \leq P_{2_\text{max}}\cdot c_0\\
0 & \leq P_{m_e} \leq P_{e_\text{max}}\cdot e_0\\
P_{\mathrm{brk}_\mathrm{min}} & \leq P_{m_\mathrm{brk}} \leq 0\\
u_3 &\in \{1,2,3\}.
\end{alignat}
\end{subequations}
This mixed-integer optimal control problem is hereinafter referred to as $\mathcal{OCP}$. Its optimal solution is calculated using the DP approach. The algorithm is an adapted version of the one presented in \cite{sundstrom2009generic} and uses a time discretization of 1 s. Both power splits use a range $u_1, u_2 \in [-1, 1]$, which allows for load-point shifting and is a range commonly found in the literature \cite{sundstrom2009generic}.\\
In this paper all solutions are coded and implemented in MATLAB\footnote{MATLAB is a registered trademark of The MathWorks, Inc.}
\subsection{Simplified Powertrain Model} \label{subsec:simplified_powertrain_model}
A simplified powertrain model, which is used to formulate an efficient MPC controller later on, is introduced in the following. A detailed description, including the validation of the simplified powertrain model, is presented in our previous work \cite{machacek2022multi}. The proposed control-oriented model uses speed-dependent convex representations of the individual drivetrain components. The required rotational speeds are computed as follows:
\begin{equation} \label{eq:rotational_speeds}
	\begin{aligned}
		\omega_{1} & = \frac{v\cdot \gamma_{1}}{r} \\
		\omega_{e} & = \frac{v\cdot \gamma_{\mathrm{diff}}\cdot \gamma_{\mathrm{GB}}(v)}{r}\cdot c_0\\
		\omega_{2} & = \omega_{e} \cdot \gamma_{2}\cdot c_0. 
	\end{aligned}
\end{equation}
The transmission ratios for the corresponding components are $\gamma_{\mathrm{GB}}, \gamma_{1}, \gamma_{2}, \gamma_{\mathrm{diff}}$ for the engine gearbox, motor 1 transmission, motor 2 transmission, and the differential transmission. The gear choice is defined by a map that depends on the vehicle speed, and the engine speed and motor 2 speed are assumed to be zero when the clutch is open.\\
\\
The fuel consumption map is approximated using a speed-dependent quadratic Willans approach. The fuel consumption is transformed to result in fuel power as follows: 
\begin{equation} \label{eq:convex_fuel}
\begin{aligned}
P_f &= \kappa_0(\omega_{e})\cdot e_0 + \kappa_1(\omega_{e}) \cdot P_{m_e} + \kappa_2(\omega_{e}) \cdot P_{m_e}^2\\
\dot{m}_f &= \frac{P_f}{H_l},
\end{aligned}
\end{equation}
where $H_l$ is the lower heating value of gasoline.\\
\\
The electric motor loss maps are approximated using a speed-dependent piecewise linear approximation as follows:
\begin{equation} \label{eq:convex_losses}
	\begin{aligned}
		P_{l_1} &= a_{1,j}(\omega_1)\cdot P_{m_1} + b_{1,j}(\omega_1)\\
		P_{l_2} &= a_{2,k}(\omega_2)\cdot P_{m_2} + b_{2,k}(\omega_2).
	\end{aligned}
\end{equation}
A good fit was achieved using ten linear functions for \mbox{motor 1} and four linear functions for \mbox{motor 2}, i.e., $j \in \{1,\ldots,10\}$ and $k \in \{1,\ldots,4\}$.\\
\\
The simplified battery model is based on a constant battery open circuit voltage $V_{oc}=46 \,\mathrm{V}$. Using a piecewise quadratic relation, it maps electric power request onto battery source power, including battery internal losses:
\begin{equation}
P_{b_s} = 
\begin{cases}
r_1^+ \cdot P_b + r_2^+ \cdot P_b^2 \quad \quad \quad &P_{b} \geq 0 \\
r_1^- \cdot P_b + r_2^- \cdot P_b^2 \quad \quad \quad &P_{b} < 0.
\end{cases} 
\label{eq:convex_battery}
\end{equation}
\\
\subsection{Drive Cycles}
\begin{figure}[h!]
	\centering
	\includegraphics[scale=1]{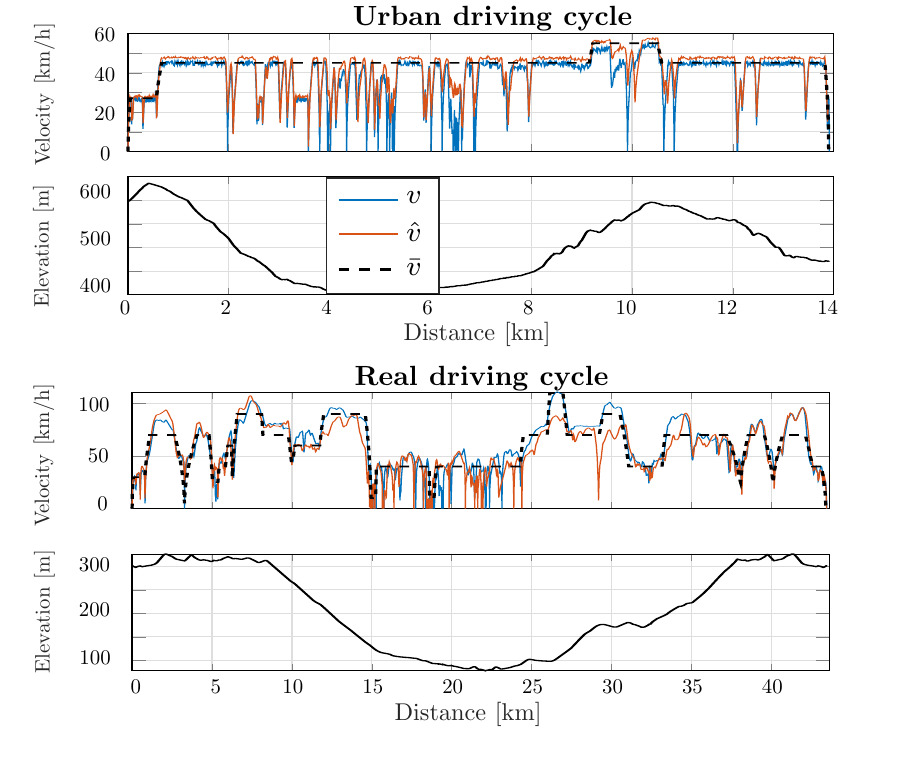}\\
	\vspace{-0.6cm}
	\caption{Blue depicts the real driving mission. Red depicts the velocity that is used as prediction. Dashed black depicts the velocity that is used for the RTG.}
	\label{fig:cycle_both}
\end{figure}

Two drive cycles will be used in the remainder of this text to showcase the optimal driving mode distribution and to assess the performance of the LB-MPC controller.
The first drive cycle is extracted from the open-source software SUMO \cite{behrisch2011sumo} and is referred to as \textit{urban cycle}. The second drive cycle is based on recorded sensor data of a real vehicle and features a mix of urban driving, rural driving, and highway driving and is referred to as \textit{real driving cycle}.\\
\\
The \textit{urban cycle} and the \textit{real driving cycle} are depicted in Figure \ref{fig:cycle_both}. Both cycles are displayed over the travel \mbox{distance $s$} and depict three velocity trajectories in the upper part and the elevation profile in the lower part. 
The blue lines depict the actual vehicle speed $v$, which is interpreted as the driver's request and is not known a priori. The red lines depict a predicted velocity $\hat{v}$. The dashed black lines depict a crude velocity estimation $\bar{v}$. Further elaborations on the difference between $v, \hat{v}, \bar{v}$ will be in Section \ref{sec:controller_structure}.

\section{Driving Mode Analysis} \label{sec:driving_mode_analysis}
In this section, the optimal choice for the driving mode is explored using optimal control theory and the optimal solution calculated using DP. The findings will be used in Section~\ref{sec:controller_structure} to develop a method that approximates the $\mathcal{OCP}$ but can be computed much quicker.

\subsection{Pontryagin’s Minimum Principle}
PMP states the necessary conditions for optimality of an optimal control problem after reformulating it as a Hamiltonian system. These can provide insight into the optimal solution without actually computing it. If the $\mathcal{OCP}$ is formulated for the simplified model equations of Section \ref{subsec:simplified_powertrain_model}, the corresponding Hamiltonian function reads:
\begin{equation} \label{eq:Hamiltonian}
	\begin{aligned}
	H(x(t),\mathbf{u}(t),\lambda(t)) &= P_f(\mathbf{u}(t)) + \lambda(t)\, \dot{E}_b(t)\\
							&= P_f(\mathbf{u}(t)) - \lambda(t)\, P_{s_b}(\mathbf{u}(t)).
	\end{aligned}
\end{equation}
The variable $\lambda$ is the costate associated to the dynamics of the system's state (\ref{eq:OCP_state}). If the Hamiltonian does not depend on the state, the costate is either constant (if the state bounds are not reached) or piecewise constant (if the bounds  are reached one or multiple times)~\cite{ritzmann2019fuel}. The optimal value(s) of the costate can be found by solving a two-point boundary value problem~\cite{diehl2016NumOpt}. However, this is neither straightforward nor needed for the derivations that follow.\\

For a given optimal trajectory of the costate, the optimal control inputs can be found by minimizing the Hamiltonian for each point in time:
\begin{equation} \label{eq:optctrlaw}
	\begin{aligned}
		\mathbf{u}^\star(t)& = \underset{\:\: \mathbf{u}\hspace*{0.1em}\in\hspace*{0.1em}\mathcal{U}(t)}{\arg\min}\: H(\mathbf{u}(t),\lambda^\star(t))\\
		&\text{s.t.} \quad P_{\text{GB}_2} = P_{\text{req}}-P_{m_1} - P_{\text{GB}_e} - P_{\text{GB}_\text{brk}}
	\end{aligned}
\end{equation}
The star in the superscripts denotes optimality. It is practically useful to carry out this minimization by separating the continuous and integer control inputs:
\begin{subequations} \label{eq:MinHcalc}
	\begin{align}
		H_i^\star(t) &= \underset{u_1, u_2}{\min}\: H(u_1, u_2,u_3 = i,\lambda^\star(t)) \label{eq:Hi}\\
		u_3^\star(t) &= \underset{\:\: u_3}{\arg\min}\: \{ H_1^\star(t), H_2^\star(t), H_3^\star(t) \} \label{eq:u3opt}\\
		\mathbf{u}^\star(t) &= \underset{u_1, u_2}{\arg\min}\: H(u_1, u_2,u_3 = u_3^\star,\lambda^\star(t))\\
		&\text{s.t.} \quad P_{\text{GB}_2} = P_{\text{req}}-P_{m_1} - P_{\text{GB}_e} - P_{\text{GB}_\text{brk}}
	\end{align}
\end{subequations}
Note that for the calculation of~\eqref{eq:Hi}, the respective formulation of the requested force from~\eqref{eq:vehiclePower} must be used.\\

\subsection{Critical Power Request} \label{subsec:critical_power_request}
\begin{figure}[h]
	\vspace{-0.6cm}
	\centering
	\includegraphics[width=0.95\columnwidth]{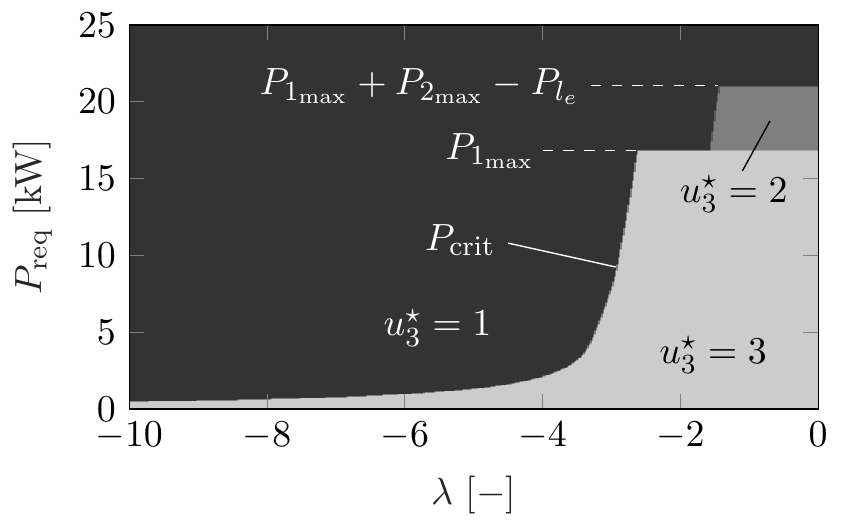}\\
	\vspace{-0.3cm}
	\caption{Optimal driving mode depending on the battery costate and requested power, for a constant vehicle speed of 50 km/h and fourth gear.}
	\label{fig:u3opt_vs_lambda_and_Preq}
\end{figure}

Figure \ref{fig:u3opt_vs_lambda_and_Preq} shows the optimal driving mode, calculated using (\ref{eq:u3opt}), depending on the battery costate and requested power, for a constant vehicle speed of 50 km/h and fourth gear. It can be seen that a distinct separation exists between HEV mode ($u_3 = 1$) and EV mode ($u_3 = 3$). For a given $\lambda$, there seems to exist a unique critical power request, $P_\mathrm{crit}$, which separates the optimal driving modes. In fact, for a simplified drivetrain configuration, an algebraic expression could be derived for $P_\mathrm{crit}$ (see Appendix~\ref{sec:appendixPreqcrit}).\\

The investigation of the necessary conditions of optimality using the Hamiltonian for the simplified drivetrain model leads to the following two theoretical findings:
\begin{enumerate}
	\item[($\mathcal{L}_1$)] The existence of a critical power request for a given vehicle speed and gear means that, for an a priori known $\lambda$, $u_3^\star$ can be determined independent of $u_1$ and $u_2$.
	\item[($\mathcal{L}_2$)] The existence of a critical power request for a given $\lambda$ means that $P_\mathrm{crit}$ is drive cycle-dependent and usually not known a priori.
\end{enumerate}

\begin{figure}[h]
	\vspace{-0.3cm}
	\centering
	\includegraphics[width=\columnwidth]{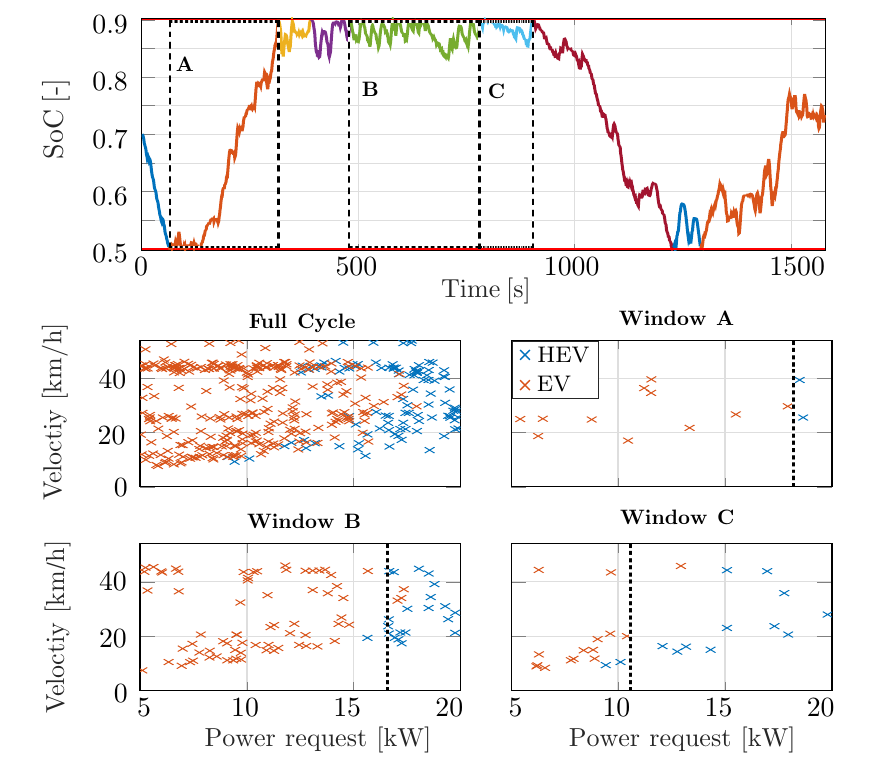}\\
	\vspace{-0.3cm}
	\caption{The top plot displays the optimal SoC trajectory, as calculated using DP, for the \textit{urban cycle}. Battery state bound activations are emphasized using different colors for the SoC trajectory, whereas three time windows are highlighted using the dashed black boxes. The resulting optimal driving mode selections for the full cycle and the time windows are shown in the corresponding subplots. Recuperation mode is only activated for negative power requests, and therefore not displayed here.}
	\label{fig:DPanalysis}
\end{figure}

While the optimal costate is (piecewise) constant when using the simplified model, this is generally not the case for the vehicle speed and gear selection. Thus, $P_\mathrm{crit}(t)$ will generally not be (piecewise) constant. Moreover, in the full model, the open circuit voltage depends on the state of charge, so that the costate is not piecewise constant. To test $\mathcal{L}_1$ and $\mathcal{L}_2$ for the full model, the solution of the $\mathcal{OCP}$ was calculated on the \textit{urban cycle} using DP. Figure \ref{fig:DPanalysis} shows in the upper part the optimal battery state trajectory. The vertical dotted lines denote battery state constraints activations, which can lead to jumps in the optimal costate~\cite{ritzmann2019fuel}. Some of these are used to split the full trajectory into the time windows A-C. The lower part displays the optimal driving mode distribution for the full driving mission and time windows A-C in the \mbox{($P_\mathrm{req},\,v$)-space}. The red crosses denote that EV mode is optimal, and the blue crosses denote that HEV mode is optimal. Indeed, whereas the optimal distribution for the full driving cycle shows a blurry boundary between EV mode and HEV mode, a clear separation exists between the optimal driving modes for each period of time during which the state constraints are not active. This shows that $\mathcal{L}_1$ and $\mathcal{L}_2$ still hold reasonably well for the full model. The finding $\mathcal{L}_1$ will be used in \mbox{Section \ref{sec:controller_structure}} to motivate an approximation for the $\mathcal{OCP}$ that can be solved efficiently within an MPC scheme. The finding $\mathcal{L}_2$ will be used in \mbox{Section \ref{sec:learning_algorithm}} to motivate a learning algorithm to estimate $P_\mathrm{crit}$ online.

\section{Controller Structure} \label{sec:controller_structure}
\begin{figure}[h!]
\centering
   \def\svgwidth{1\columnwidth}
   {\scriptsize
	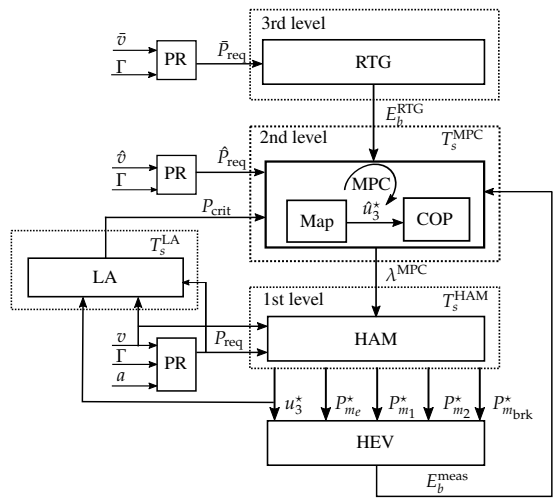}
\caption{Schematic illustration of the multi-level MPC controller structure. On the 1st level, the control inputs are calculated by the HAM based on an estimation of $\lambda^\mathrm{MPC}$. On the 2nd level, the MPC estimates the optimal battery costate based on a prediction of the driving mission and a battery reference trajectory. On the 3rd level, the RTG calculates the battery reference trajectory for the full driving mission.}
\label{fig:controller_structure}
\end{figure}

In this section, the control-levels are explained using a bottom-up-approach and the controller level's inputs and outputs are highlighted. Figure \ref{fig:controller_structure} depicts the proposed multi-level LB-MPC structure. All levels receive a power request as input, which is denoted by the PR-block. It calculates the power request using (\ref{eq:vehiclePower}), (\ref{eq:force_to_power}), and the corresponding signals for the velocity, the elevation, and possibly the acceleration (refer to Figure \ref{fig:cycle_both}).
\subsection{Hamiltonian}
The first control level corresponds to a static optimization of the Hamiltonian (\ref{eq:optctrlaw}) and is used to calculate the desired controller inputs in real time. It is abbreviated with HAM within this text. 
Using an input grid over all controller inputs $\mathbf{u}$ and the power demand of the driver, as well as the measured velocity $v$, the Hamiltonian is minimized as in (\ref{eq:MinHcalc}). Assuming that an estimate of the costate $\lambda$ is available, the solution to this static optimization can be calculated very efficiently.
\subsection{MPC}
The second control level is based on a model predictive control approach to compute an estimate of the costate trajectory $\lambda^\mathrm{MPC}$ every $T^\mathrm{MPC}$ seconds. Since it is not viable to solve the $\mathcal{OCP}$ in real-time, an approximation based on $\mathcal{L}_1$ is solved instead. Assuming that the cycle-dependent $P_\mathrm{crit}$ is known, the predicted power request can be used to pre-compute $\hat{u}^\star_3$. The following control map is encoded in the Map-block:
\begin{equation}
	\begin{aligned}
		\hat{u}^\star_3 = 1, \quad &\hat{P}_\mathrm{req} \geq P_\mathrm{req}^\mathrm{crit}\\
		\hat{u}^\star_3 = 2, \quad &\hat{P}_\mathrm{req} < (P_{1_\mathrm{min}}-P_{l_e})\\
		\hat{u}^\star_3 = 3, \quad &(P_{1_\mathrm{min}}-P_{l_e}) < \hat{P}_\mathrm{req} < P_\mathrm{req}^\mathrm{crit},
	\end{aligned}
\end{equation}
where $\hat{P}_\mathrm{req}$ is calculated using (\ref{eq:vehiclePower}) and (\ref{eq:force_to_power}), while assuming HEV mode for the full prediction. 
Since the Recuperation mode leads to substantial engine friction losses, this mode is only chosen if the power that could be recuperated exceeds $P_{1_\mathrm{min}}-P_{l_e}$, which then allows for further recuperation with the help of motor 2. Using Recuperation mode for positive power requests only makes sense in driving scenarios which provide excessive downhill sections, and are not considered in this work.\\
\\
With the use of a pre-computed $\hat{u}^\star_3$ and the simplified model, the $\mathcal{OCP}$ can be reformulated as a convex optimal control problem (COP). Constraint relaxations for (\ref{eq:convex_fuel}), (\ref{eq:convex_battery}), and (\ref{eq:convex_losses}) are used to describe the feasible solution-set as a convex set. The constraint relaxations are based on \cite{murgovski2015convex} and are described in detail in our previous work \cite{machacek2022multi}. For the sake of readability all motor speeds and time dependencies are omitted in the following. Furthermore, a reformulated vector of control inputs $\mathbf{\tilde{u}}=[P_{m_1}, P_{\mathrm{GB}_e}, P_{\mathrm{GB}_\mathrm{brk}}]$ is used, which directly follows the equations describing the power splits in Section \ref{subsec:powertrain_model}. The time discretization is 1s.
\begin{subequations}\label{eq:COP}
	\begin{alignat}{2}
	    \underset{\tilde{\mathbf{u}}}{\text{min}} & \int_{t_0}^{t_0+H_p}  P_f \, dt \label{eq:COP_cost}\\
		\text{   \qquad s.t.} \quad \dot{E_b}(t) & = -P_{s_b} \label{eq:COP_statedynamics}\\
		E_b(t_0) & = E_{b_0}\\
		E_b(t_0+H_p) & \geq E_b^\mathrm{RTG} \label{eq:OCP_terminal} \\
		E_{b_\mathrm{min}} & \leq E_b(t) \leq E_{b_\mathrm{max}} \label{eq:OCP_state_bounds} \\
		P_{\text{GB}_2} & = \hat{P}_{\text{req}}-P_{m_1} - P_{\text{GB}_e} - P_{\text{GB}_\text{brk}} \\
		P_{m_2} & \geq P_{\text{GB}_2} \cdot \eta_{\mathrm{GB}} \label{eq:OCP_simplified_sup1} \\
		P_{m_2} & \geq \nicefrac{P_{\text{GB}_2}}{\eta_{\mathrm{GB}}} \label{eq:OCP_simplified_sup2} \\
		P_{l_1} & \geq a_{1,j} P_{m_1} + b_{1,j} \label{eq:OCP_simplified_sup3} \\
		P_{l_2} & \geq a_{2,k} P_{m_2} + b_{2,k} \label{eq:OCP_simplified_sup4} \\
		P_{s_b} & \geq r_1^+ P_b + r_2^+ P_b^2 \label{eq:OCP_simplified_sup5} \\
		P_{s_b} & \geq r_1^- P_b + r_2^- P_b^2 \label{eq:OCP_simplified_sup6} \\
		P_f & \geq \hat{e}_0 \cdot \kappa_0 + \kappa_1 \frac{P_{\mathrm{GB}_e}}{\eta_{\mathrm{GB}}} + \kappa_2 \left(\frac{P_{\mathrm{GB}_e}}{\eta_{\mathrm{GB}}}\right)^2	\label{eq:OCP_simplified_sup7} \\
		0 & \leq \nicefrac{P_{\mathrm{GB}_e}}{\eta_{\mathrm{GB}}} \leq \hat{e}_0 \cdot P_{e_\text{max}} \label{eq:OCP_state_pe} \\
		P_{1_\text{min}} & \leq P_{m_1} \leq P_{1_\text{max}}\\
		\hat{c}_0 \cdot P_{2_\text{min}} & \leq P_{m_2} \leq  \hat{c}_0 \cdot P_{2_\text{max}} \label{eq:OCP_state_p2} \\
		P_{\mathrm{brk}_\mathrm{min}} & \leq P_{m_\mathrm{brk}} \leq 0
	\end{alignat}
\end{subequations}
The MPC has a prediction horizon $H_p$. At each update, it receives a predicted power request trajectory $\hat{P}_\mathrm{req}$ and a battery energy target $E_b^\mathrm{RTG}$ that is used as terminal constraint.\\
\\
This COP is parsed using YALMIP \cite{lofberg2004yalmip}. The solver MOSEK \cite{mosek} is used to solve the COP and takes roughly 1 s to compute the solution for a 30 min drive cycle on a standard laptop (quadcore processor \mbox{@ 3.3 GHz}).
\subsection{Reference Trajectory Generator}
The third control level is used to calculate a suitable battery reference trajectory and is referred to as RTG within this text. The battery reference trajectory  $E_b^\mathrm{RTG}$, which is used as target trajectory by the underlying MPC, is calculated by solving the COP prior to departure for the entire driving mission. The predicted power request $\bar{P}_\mathrm{req}$ is calculated using the elevation profile and a crude velocity prediction $\bar{v}$, which corresponds to the speeding limits of the driving mission. On this controller level, the driving mode is assumed to be always HEV mode.
\section{Learning Algorithm} \label{sec:learning_algorithm}
The finding $\mathcal{L}_2$ has shown that $P_\mathrm{crit}$ is cycle-dependent and can jump if the battery state constraints are activated. Therefore, it is desirable to adapt the control map during the driving mission. The LA is used to learn the parameter $P_\mathrm{crit}$.
\subsection{Classification Problem}
\begin{figure}[h!]
\centering
   \def\svgwidth{1\columnwidth}
   {\scriptsize
	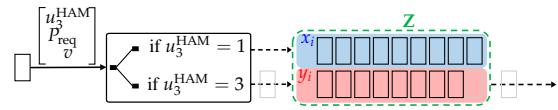}\\
	\vspace{-0.3cm}
	\caption{Schematic of how the buffers are filled.}
	\label{fig:buffers}
\end{figure}

While the vehicle is driving, pairs of $z = (P_\mathrm{req}, v)$ are stored in a buffer with length $n_B$ using the first-in first-out principle. Figure \ref{fig:buffers} depicts how they are distributed based on the driving mode selection of the HAM $u_3^\mathrm{HAM}$. Here, $\mathbf{Z}$ denotes the set of all pairs $z_i,$ for $i\in \{1,\ldots,n_B\}$, $x_i$ denotes a pair that corresponds to $u_3^\mathrm{HAM}=1$, and $y_i$ denotes a pair that corresponds to $u_3^\mathrm{HAM}=3$.
\\
The information in the buffer is used to learn the decision of the HAM regarding the optimal driving mode based on the costate $\lambda^\mathrm{MPC}$. Since $P_\mathrm{crit}$ separates the regions \mbox{$u_3=1$} and \mbox{$u_3=3$} in the ($P_\mathrm{req},\,v$)-space, an affine function \mbox{$f(z) = \mathbf{m}^Tz-q$} can be fitted to separate the classified points $x_i, y_i$ as follows: 
\begin{equation}
	\begin{aligned}
		\mathbf{m}^Tx_i-q&>0,\\
		\mathbf{m}^Ty_i-q&<0.
	\end{aligned}
\label{eq:discrimination}
\end{equation}
Such a classification problem can be written as a quadratic program (QP):
\begin{subequations}\label{eq:lp}
	\begin{alignat}{2}
		\underset{\mathbf{m},q,r_1, r_2}{\text{min}} \quad   ||m||_2 + &\gamma(\mathbf{1}^Tr_1 + \mathbf{1}^Tr_2)\\
		\text{  \qquad s.t.} \quad m^Tx_i - q  & \geq 1 - r_1\\
		\quad m^Ty_i - q & \leq -(1-r_2)\\
		r_1  \geq 0, \, r_2 & \geq 0\\
		m(1) \geq 0, \, m(2) & \geq 0 \label{eq:lp1}\\
		m(1) & \geq 10^5\cdot m(2) \label{eq:lp2}\\
		0\leq q & \leq 10^4 \label{eq:lp3}
	\end{alignat}
\end{subequations}
The classification problem is slacked using support vectors to ensure the feasibility of the QP even if the groups aren't strictly separable. If there exists an affine function that strictly classifies the groups, the optimal cost will be reached with $r_1=r_2=0$ and the non-slacked constraints are recovered. If slacking is required, then the parameter $\gamma>0$ serves as a tuning parameter that gives the relative weight of the number of misclassified points, compared to the width of a slab ($\propto \nicefrac{1}{||m||_2}$) that is drawn between the two groups \cite{boyd2004convex}. Because a constant $P_\mathrm{crit}$ separates the optimal driving modes, (\ref{eq:lp1}) and (\ref{eq:lp2}) enforce a very large and positive slope of the affine function. Equation (\ref{eq:lp3}) increases the numerical stability of the QP by limiting the feasible set for $q$ to a reasonable range. The learning process is updated periodically every $T_s^\mathrm{LA}$ seconds using the buffered data to compute a new solution to the QP. The critical power is then calculated as
\begin{equation}
	P_\mathrm{crit} =
	\begin{cases}
		\frac{q}{m(1)} & \lambda > 0,\\
		P_1^\mathrm{max} & \lambda = 0.
	\end{cases}
\label{eq:p_crit}
\end{equation}
The variable $P_\mathrm{crit} = P_1^\mathrm{max}$ is set, if $\lambda^\mathrm{MPC}=0$ is detected for at least nine consecutive seconds. This is done because $\lambda^\mathrm{MPC}=0$ indicates that the equivalent fuel consumption of electric energy is zero, i.e., EV mode should always be favoured if possible to use as much electric energy as possible. This behavior ensues if the predicted driving mission for the MPC's prediction horizon offers more energy to recuperate than is necessary to fulfill its terminal battery constraint.\\
Overall, the LA serves as a direct feedback-loop from the control decisions of the HAM to the MPC. Thanks to this, the MPC receives information on the effect of $\lambda^\mathrm{MPC}$ on the additional degree of freedom $u_3^\mathrm{HAM}$.
\subsection{Learning Performance}
\begin{figure}[h]
	\centering
	\includegraphics[width=\columnwidth]{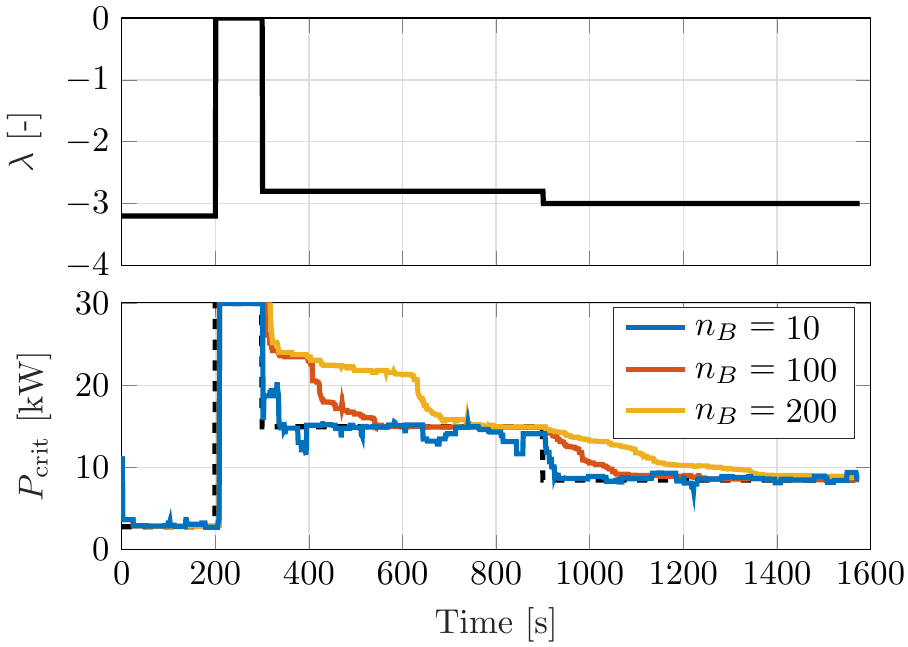}\\
	\vspace{-0.3cm}
	\caption{Performance of the LA for various buffer sizes. The dashed black line in the lower plot shows $P_\mathrm{crit}$, which results from the $\lambda$-trajectory in the upper plot and (\ref{eq:u3opt}).}
	\label{fig:case_study_learning_algorithm}
\end{figure}

In this section, the LA is analyzed and the effect of the buffer size on the convergence is discussed. To focus on the learning performance, disturbances occurring due to model mismatch or mispredicted velocity data, are eliminated by simulating the LA-block only (refer to Figure \ref{fig:controller_structure}). The LA receives pairs of $(P_\mathrm{req},v)$. They are stored in its buffer according to their respective driving mode $u_3$, which is calculated using (\ref{eq:u3opt}), a predefined $\lambda$ trajectory, and a constant vehicle speed of 50 km/h and fourth gear (refer to Figure \ref{fig:u3opt_vs_lambda_and_Preq}). The data is gathered and stored in the buffer at a frequency of 1 Hz and the QP is solved every 10 s.\\
\\
Figure \ref{fig:case_study_learning_algorithm} shows in the upper plot the piecewise constant costate trajectory that is used to showcase the LA's learning performance. The lower plot displays $P_\mathrm{crit}$, which follows from (\ref{eq:u3opt}), with the dashed black line, and the performance of the LA for various values of $n_B$ in blue, red, and orange, respectively. 
The buffer size is a tuning parameter, which has a direct impact on the convergence speed of the LA after a change of $P_\mathrm{crit}$. The larger $n_B$, the longer old $z_i$ are stored. This can be seen by the red and yellow curves in the time windows \mbox{$t\in$ [300 s, 600 s]} and \mbox{$t\in$ [900 s, 1150 s]}, where old pairs of $z_i$, which correspond to an old $P_\mathrm{crit}$, are still in the buffer. During these time windows, the $z_i$ are not fully separable and hence, the QP needs slacking and only converges to the true value after the old $z_i$ are deleted from the buffer. A small $n_B$ results in a faster convergence after a change of $P_\mathrm{crit}$. However, this also means that important $z_i$ are forgotten quickly, which can lead to oscillating behavior around the optimum. This can be seen in the blue curve during \mbox{$t\in$ [610 s, 900 s]}. The update frequency $T_s^\mathrm{LA}$ only has a minor effect on the result and was not changed anymore.

\section{Implementation for Online Control} \label{sec:online_implementation}
All presented controllers are implemented in the time domain, which needs to be taken into account when updating the predictive data. Prior to each MPC update, the measured vehicle position $s(t)$ is matched with the previously predicted vehicle position and the predictive data is shifted accordingly. Although, updated predictive data could be used at every re-initialization of the MPC, the same predictions are used throughout this case study. It is assumed that updating the velocity predictions according to newer traffic data should even improve the prediction quality.\\
Without further restriction, it is assumed that the predicted $\hat{P}_\mathrm{req}$ does not surpass the maximum power of the entire drivetrain at any point in time.\\
\\
A major concern regarding the use of model predictive control is to ensure the feasibility of the underlying optimal control problem. In a real-case scenario, the existence of a feasible solution to the COP - and therefore the viability of the LB-MPC in general - is threatened by different limiting factors. The following challenges are looked at in particular:\\
\begin{enumerate}
	\item Standard state-of-the-art onboard tools only provide detailed velocity predictions for a limited look-ahead horizon. A full prediction horizon is unrealistic and the MPC must be formulated using a receding horizon.
	\item The computation time of the MPC should be well below its update time $T_s^\mathrm{MPC}$ to ensure that at each controller iteration a new solution is at hand.
	\item The LB-MPC is subject to velocity mispredictions $\hat{v}$ and therefore also false estimations of $\hat{u}^\star_3$. The controller has to be robust to mispredictions, which will be highlighted in the following case-studies.\\
\end{enumerate}
A common technique to increase the robustness of a model predictive controller includes softening the state constraints. However, the classical approach of using soft constraints to slack the state bounds (\ref{eq:OCP_state_bounds}), and/or to slack the terminal state condition (\ref{eq:OCP_terminal}) has a direct effect on the costate $\lambda^\mathrm{MPC}$ (see Appendix \ref{sec:appendixLambdaEb}). Since this costate greatly impacts the final choice of controller inputs in the HAM, this conventional method of state constraint slacking is undesirable in this case. The following reformulations of (\ref{eq:COP_cost}) and (\ref{eq:OCP_state_pe}) are proposed, which ensure the feasibility of the convex optimization problem:
\begin{equation}
	\begin{aligned}
	    \underset{\tilde{\mathbf{u}},\,\epsilon_\mathrm{Pe}}{\text{min}} \int_{t_0}^{t_0+H_p}  (P_f + 10^5\cdot & \epsilon_\mathrm{Pe}) \, dt\\
		\nicefrac{P_{\mathrm{GB}_e}}{\eta_{\mathrm{GB}}} &\leq P_{e_\text{max}}\cdot c_0 + \epsilon_\mathrm{Pe}\\
		\epsilon_\mathrm{Pe} &\geq 0.
	\end{aligned}
\end{equation}
Two things will be highlighted in the following. First, the slacking variable $\epsilon_\mathrm{P_e} \in \mathcal{R}^{H_p\times 1}$ can be understood as a slacking of $\hat{u}^\star_3$, which is an exogenous variable that the COP cannot change directly. If a poorly predicted $\hat{P}_\mathrm{req}$, or $P_\mathrm{crit}$ close to the lower battery constraint force the COP to use EV mode ($c_0=0$), then an infeasible problem formulation can ensue. The slacking variable $\epsilon_\mathrm{Pe}$ is a tool to provide the COP with engine power if needed to ensure the feasibility of the optimization problem. It is important to ensure that the constraint is only slacked if otherwise no feasible solution exists. The reformulated cost function, together with the non-negativity constraint on $\epsilon_\mathrm{P_e}$, ensure that the non-slacked solution to the COP is recovered if it exists. The weighting of $\epsilon_\mathrm{P_e}$ has to be at least as high as the optimal Lagrange multiplier for the original problem \cite{kerrigan2000soft}. Since this Lagrange multiplier is not available beforehand, the weighting is simply chosen large enough, i.e. $10^5$ here.\\
\\
Second, if slacking is required to ensure the COP's feasibility, it is used as conservatively as possible because $\epsilon_\mathrm{Pe}$ can only increase the total cost of the optimization problem. Therefore, the timing and the amplitude of the slacking variable are chosen optimally by the COP to minimize the accumulated cost. The amplitude of the slacking variable corresponds to the additional amount of engine power that is required. Therefore, if a slacked solution is detected (i.e., $any(\epsilon_\mathrm{P_e}(t_0:t_0+H_p)>0$)) then $\hat{u}^\star_3$ is changed at these instances from EV mode to HEV mode and the COP is reiterated, effectively solving a slightly adapted optimization problem, before passing $\lambda^\mathrm{MPC}$ on to the HAM. Since the predicted $\hat{P}_\mathrm{req}$ doesn't surpass the maximum power of the entire drivetrain, the reiterated solution of the COP will always be feasible without the need of slacking any constraints and therefore, $\lambda^\mathrm{MPC}$ is not effected anymore. 

\section{Case Study} \label{sec:case_study}
In this section, the performance of the LB-MPC is assessed in simulation by comparing its results to a DP benchmark solution. The full model is used to simulate the vehicle. Based on the assumption that the route is known in advance (e.g using a navigation system), perfect knowledge of the spatial discretization of the elevation profile is assumed on all controller levels. However, special attention has to be paid to the controller's robustness to velocity mispredictions, as without a cruise controller, and in the presence of other road users, the vehicle speed can only be estimated and perfect prediction is not possible. Different representations of velocity mispredictions are used to emulate physically meaningful predictive data. They are denoted with $v, \hat{v}, \bar{v}$, and are shown graphically in Figure \ref{fig:cycle_both}.\\
\begin{itemize}
	\item The velocity $\bar{v}$ is assumed to be known for the entire driving mission and used in the RTG to calculate the battery reference trajectory. The accuracy of this signal could represent the legal speeding limits for the route.
	\item The velocity $\hat{v}$ is assumed to be known for a limited look-ahead horizon of 450 s and used in the MPC. A horizon \mbox{$H_p=450$ s} equals 7.5 km when driving with 60 km/h. This signal was generated, either by simulating the driving mission twice with different levels of traffic congestion (\textit{urban cycle}), or by taking an independent velocity measurement of the same vehicle that drove on the same route a second time (\textit{real driving cycle}).
	\item The velocity $v$ is the driver's desired velocity and therefore an unknown signal to both predictive control layers. It is used in the HAM to calculate the powersplit.\\
\end{itemize}
First, two variations of the LB-MPC are presented to serve as a comparison to highlight the effects of using a control map to estimate $\hat{u}^\star_3$ at all, and of additionally adapting this map online using the LA. Second, the performance of these two comparison controllers and the LB-MPC are evaluated on two different drive cycles.
\subsection{Comparison Controllers}
As baseline controller, which does not exploit $\mathcal{L}_1$ and $\mathcal{L}_2$, the following variation of the presented LB-MPC is proposed. The model predictive controller on the second control layer calculates $\lambda^\mathrm{MPC}$ without an estimate $\hat{u}^\star_3$ that stems from a control map. But since $\hat{u}^\star_3$ has to be predefined in order to resolve the issue of otherwise facing a mixed-integer optimization, the MPC assumes that the vehicle is always operated in HEV mode. This gives the COP access to the drivetrain's full propulsion power at all times and therefore ensures its feasibility. The costate is subsequently used in the HAM to calculate the power split, as well as the driving mode based on the momentary power request of the driver. This controller will be referred to as "baseline MPC".\\
\\
A more advanced controller variation adopts the idea of recent publications (e.g. (\cite{kwon2022control}, or \cite{mashadi2020fuel})) that suggest the use of DP-optimizations to create offline pre-calculated control maps to separate the optimal driving modes in the ($P_\mathrm{req},\, v$)-space. DP was used to calculate the optimal solution of the $\mathcal{OCP}$ on the well-known NEDC cycle. Since on this driving mission no battery constraints are activated, a nearly optimal driving mode separation is achieved with $P_\mathrm{crit}\approx\, 800\, \mathrm{W}$. Omitting the LA of the LB-MPC and always estimating $\hat{u}^\star_3$ based on $P_\mathrm{crit} = 800\, \mathrm{W}$, is equivalent to using a predefined control map to choose the driving mode for the COP. This controller will be referred to as "map-based MPC" and can be used to later highlight the importance of estimating the cycle-specific $P_\mathrm{crit}$ during online control. 
\subsection{Results}
\begin{figure}[h!]
\centering
   \def\svgwidth{1\columnwidth}
   {\scriptsize
	\import{pics/}{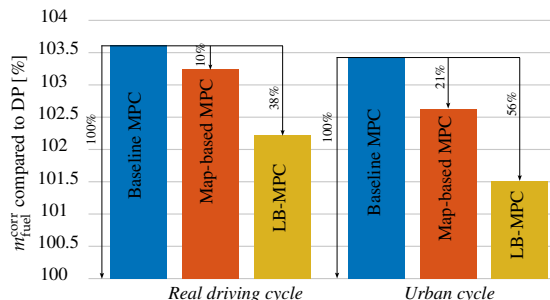}}\\
	\vspace{-0.3cm}
	\caption{Corrected fuel consumptions of baseline MPC, map-based MPC, and LB-MPC. The percentage values denote the degree of sub-optimality of the corresponding control method compared to the DP-optimal result.}
	\label{fig:results}
\end{figure}

\begin{figure}[h!]
	\centering
	\includegraphics[width=\columnwidth]{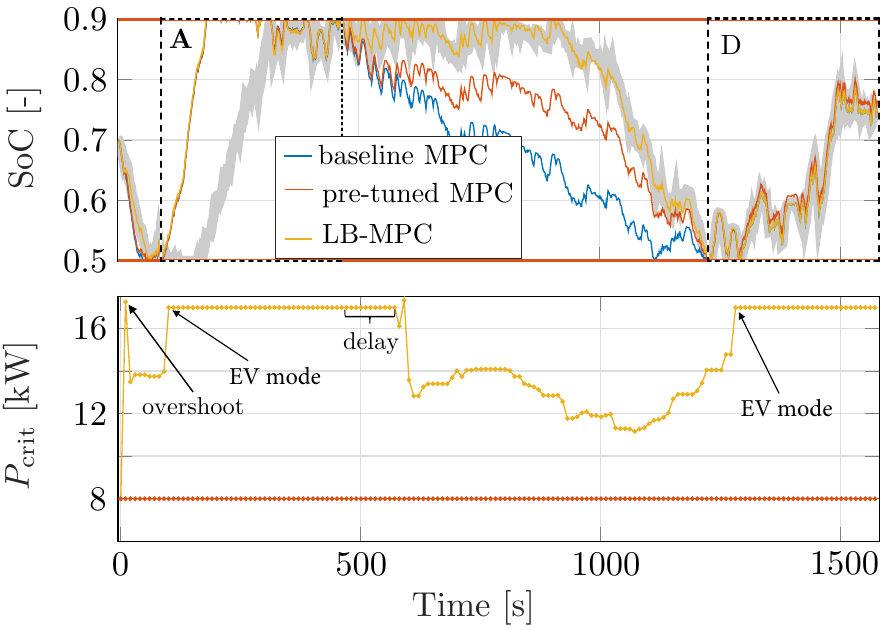}\\
	\vspace{-0.3cm}
	\caption{Top plot: Battery trajectory evolutions of the corresponding control methods on the \textit{urban cycle}. The grey line corresponds to the solution obtained with DP. Bottom plot: Values for $P_\mathrm{crit}$ of the LB-MPC and the map-based MPC, respectively. The dots represent the updated values with the update frequency that was used ($T_s^\mathrm{LA}$= 10 s).}
	\label{fig:SoC_zh}
\end{figure}

The controllers' tuning parameters are given in \mbox{Table \ref{tab:controller_parameters}} and are chosen the same for all three control methods and both driving missions. The LB-MPC is initialized with \mbox{$P_\mathrm{crit}=800$ W} and the bin size is set to $n_B=100$. Although, a faster adaption of $P_\mathrm{crit}$ is achieved by using a smaller bin size, the overall controller performance was found to be better with the slower LA.\\
\\
The corrected fuel consumption of the corresponding controller methods is used as a performance measure. The corrected fuel consumption is calculated as 
\begin{equation} \label{eq:mfuel_corr}
	m_\mathrm{fuel}^\mathrm{corr} = m_\mathrm{fuel} + \frac{1}{n} \sum_{i=1}^{n}\lambda_i^{\mathrm{MPC}} \cdot (E_b(t_f)-E_b(t_0))
\end{equation}
and accounts for the deviation of the battery energy from the final constraint. The difference in electric energy is weighted using the mean of the costates from each update of the corresponding MPC. Note that, since all three controllers are predictive, they achieve charge-sustaining vehicle operation within $\pm 2\%$.\\
\\
Figure \ref{fig:results} shows the corrected fuel consumptions of all three controller methods for both driving missions compared to the DP-optimal result in percentage, whereas 100\% is denoted by the DP solution. The map-based MPC performs better than the baseline controller for both scenarios. The \mbox{LB-MPC} learns the driving mission dependent $P_\mathrm{crit}$, effectively improving the estimation of $\hat{u}^\star_3$, and decreases the fuel consumption further. Overall, up to 56\% of the suboptimality of the baseline controller is recovered on the \textit{urban cycle}, which results in a relative performance of 101.5\% compared to the DP-optimal result. Thus, the LB-MPC achieves close-to-optimal performance.
\\
Since the fuel consumption of an HEV is tightly coupled to the battery energy allocation during the driving mission, the advantage of the LB-MPC over the two comparison controllers can be analyzed by comparing the controllers' corresponding SoC trajectories. In the upper plot of \mbox{Figure \ref{fig:SoC_zh}}, the battery trajectory evolutions of the baseline MPC, the map-based MPC, and the LB-MPC are displayed for the \textit{urban cycle} in blue, red, and yellow, respectively. The DP-optimal result is shown in grey, and is used as a comparison of how the other controllers conserve the vehicle's electric energy storage. Two time windows are highlighted with dashed boxes. In the lower plot, the values for $P_\mathrm{crit}$ are displayed for the LB-MPC and the map-based MPC.
\\
In the time windows A and D all three controller methods lead to a battery trajectory evolution that resembles the DP solution. Here, the driving mission features a distinct elevation profile, which largely determines the optimal battery state evolution. In time window A, the downhill section offers more energy than can be stored in the battery. This means that the friction brakes need to be used and different battery charging strategies can result in the same fuel consumption. In time window D, the downhill section offers more recuperable energy than is needed to fulfill the charge sustainability constraint. Therefore, the three controllers differ mostly in operating the vehicle during the time between these two windows. The operations with the baseline MPC, as well as the map-based MPC lead to a decrease in battery energy during this time, which is explained by the predefined (or not used) control map resulting in a bad estimation of $\hat{u}_3$ and subsequently an underestimation of the battery costate. The conservation of the battery energy of the LB-MPC, however, leads to a state evolution that resembles the DP solution closely.\\
\\
This driving mission features intermittent subsections with abundant recuperable energy and subsections where the rationing of the electric energy is important, which results in multiple SoC-bound activations, as well as a frequent change of $P_\mathrm{crit}$. The LA is initialized with the same critical power request as the map-based MPC, i.e. $P_\mathrm{crit}$ = 8 kW. However, it recognizes that this value is too low and adjusts it accordingly. Although the LA overshoots considerably after the first iteration, it already reaches a steady-state value of $P_\mathrm{crit} \approx$ 13.9 kW after the second learning update. The overshoot is explained by the few buffered data points at the initialization of the LA, which therefore have a large impact on its solution. When the lower SoC bound is activated at the beginning of window A, the abundant recuperable energy is reflected in the LA by setting $P_\mathrm{crit} = P_{1_\mathrm{max}}$. This means that, whenever the predicted power request can be delivered solely by motor 1, EV mode will be selected in the control map.\\
At the end of window A, the LA has a delay of 100 s until $P_\mathrm{crit}$ is updated, which reflects the buffer size $n_B=100$. From \mbox{600 s} on, the LA updates $P_\mathrm{crit}$ regularly, until EV mode is favoured again starting from \mbox{1250 s} until the end of the driving mission.

\begin{table}[h!]
\centering
\caption{Controller parameters}
\label{tab:controller_parameters}
\begin{tabular}{c| l| c| l| c| c| c}
 & $H_p$ & $T_s$ & $T_s^\mathrm{LA}$ & $n_B$ & re-iterate & $P_\mathrm{crit}$\\\hhline{-|-|-|-|-|-|-}
 \textbf{Baseline MPC} & 450 s & 2 s &  - & - & yes & -\\ 
\textbf{Map-based MPC} & 450 s & 2 s &  - & - & yes & 800 W\\ 
\textbf{LB-MPC}& 450 s & 2 s &  10 s & 100 & yes & variable \\\hline
\end{tabular}
\end{table}

\section{Conclusion} \label{sec:outlook}
In this paper, an online feasible LB-MPC is presented to tackle the EMS of HEVs including the driving modes: HEV mode, EV mode, and Recuperation mode. The structure of the control algorithm is based on the theoretical derivation of a piecewise constant, drive cycle-dependent, critical power request $P_\mathrm{crit}$ that optimally separates the driving modes. The existence of a critical power request was proven for a simplified HEV drivetrain configuration and shown to hold reasonably well for the full drivetrain model. The use of $P_\mathrm{crit}$ allows for two things:
\begin{enumerate}
	\item Formulation of a control map in the $(P_\mathrm{req},\, v)$-space to estimate the optimal driving mode $u_3^\star$. The control map is used to formulate the model predictive controller based on convex optimization theory, and ultimately, the resulting LB-MPC as a multi-level controller.
	\item Formulation of a learning algorithm to iteratively learn the drive cycle-dependent $P_\mathrm{crit}$ to update the control map during online control.
\end{enumerate}
The performance of the LB-MPC controller was compared against a baseline controller and a map-based controller, which is inspired from the literature, in two realistic driving scenarios. The proposed controller structure shows good robustness to velocity mispredictions, resulting from differing traffic conditions and the uncertainty of the vehicle's driver. In both driving scenarios, the LB-MPC achieves close-to-optimal performance.\\
\\
Future research could focus on the dynamic effects of mode switches. This could include an analysis of the time that is needed for mode switches and the implementation of the switching dynamics in the online-capable controller structure.
\setcounter{figure}{0}
\appendix
\subsection{Critical Power Request} \label{sec:appendixPreqcrit}
\begin{figure}[h]
	\centering
	\includegraphics[width=0.95\columnwidth]{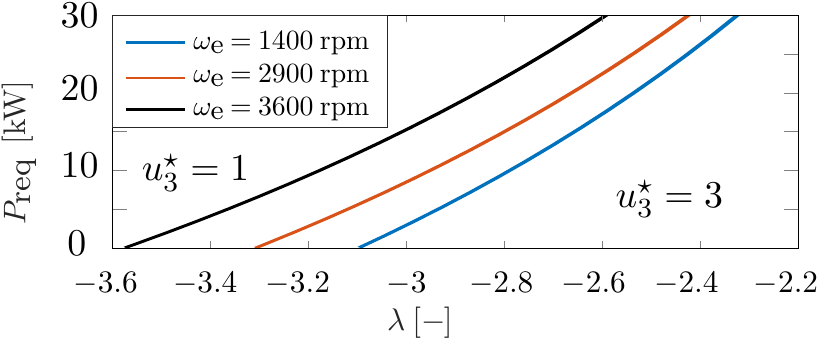}
	\caption{Separation between the HEV mode ($u_3^\star = 1$) and EV mode ($u_3^\star = 3$) by a critical power request, for a selection of rotational speeds.}
	\label{fig:Preqcrit_vs_lambda}
\end{figure}

An algebraic expression for the critical power request can be derived based on the following assumptions:
\begin{itemize}
	\item The drivetrain consists of an engine and one electric motor which is mounted somewhere between the clutch and the wheels, i.e., at position P2, P3, or P4 in a parallel hybrid configuration.
	\item The electric motor loss $P_l$ and auxiliary power $P_\mathrm{aux}$ are neglected.
	\item The gear choice is given, such that all transmission ratios are known.
	\item The vehicle drives at a constant speed, such that the engine and motors speeds are known.
	\item The gearbox efficiency is assumed to be 100 \%.
\end{itemize}
With these assumptions, the powertrain model reduces to
\begin{subequations}
\begin{align}
	P_f &= \kappa_0 + \kappa_1\, P_\mathrm{e} + \kappa_2\, P_\mathrm{e}^2, \label{eq:Pf_simple}\\
	P_\mathrm{e} &= P_\mathrm{req}\, (1-u), \label{eq:Pe_simple}\\
	P_{s_b} &= r_1\, P_m + r_2\, P_m^2, \label{eq:Psb_simple}\\
	P_m &= P_\mathrm{req}\, u, \label{eq:Pm_simple}
\end{align}
\end{subequations}
where $u$ denotes the power split between the engine and electric motor. The minimization of the Hamiltonian is carried out as in~\eqref{eq:MinHcalc}. By combining~\eqref{eq:Pf_simple}--\eqref{eq:Pm_simple}, the Hamiltonian for the HEV driving mode can be written as
\begin{subequations}
\begin{align}
	H_1 &= H(u, u_3 = 1,\lambda) = P_f(u) - \lambda\, P_{s_b}(u)\\
	&= a\, u^2 + b\, u + c \label{eq:ABC}
\end{align}
\end{subequations}
with
\begin{align}
	a &= (\kappa_2 - \lambda\, r_2)\, P_\mathrm{req}^2, \label{eq:A} \\
	b &= -( 2\, \kappa_2\, P_\mathrm{req} + \kappa_1 + \lambda\, r_1 )\, P_\mathrm{req},\\
	c &= \kappa_2\, P_\mathrm{req}^2 + \kappa_1\, P_\mathrm{req} + \kappa_0. \label{eq:C}
\end{align}
Minimizing the quadratic function~\eqref{eq:ABC} yields the minimum value of the Hamiltonian:
\begin{equation}
	H_1^\star = c - \frac{b^2}{4 a}. \label{eq:H1star}
\end{equation}
For the considered drivetrain configuration, the Recuperation mode (see Table~\ref{tab:driving_modes}) is not used, as the engine and electric motor can always be mechanically separated. For the EV mode, the Hamiltonian reads
\begin{subequations}
	\begin{align}
		H_3 &= H(u, u_3 = 3,\lambda) = - \lambda\, P_{s_b}(u)\\
		&= -\lambda\, (r_1\, P_\mathrm{req}\, u + r_2\, P_\mathrm{req}^2\, u^2).
	\end{align}
\end{subequations}
Since in this case the engine is switched off, $u = 1$, and the minimum value of the Hamiltonian equals
\begin{equation}
	H_3^\star = -\lambda\, (r_1\, P_\mathrm{req} + r_2\, P_\mathrm{req}^2).
\end{equation}
The optimal driving mode follows from~\eqref{eq:u3opt}, in which the Hamiltonian with the smallest value is sought. The critical power request $P_\mathrm{crit}$, which determines the switching point for the optimal driving mode, can thus be found by solving
\begin{equation}
	H_1^\star(P_\mathrm{crit}) \overset{!}{=} H_3^\star(P_\mathrm{crit}). \label{eq:H1eqH3}
\end{equation}
This is a quadratic equation, since plugging~\eqref{eq:A}--\eqref{eq:C} into~\eqref{eq:H1star} yields a quadratic function. Therefore, the solution to~\eqref{eq:H1eqH3} can be found using the quadratic formula:
\begin{equation}
	P_\mathrm{crit} = \frac{-\tilde{b} + \sqrt{\tilde{b}^2 - 4\, \tilde{a}\, \tilde{c}}}{2\, \tilde{a}}
\end{equation}
with
\begin{align}
	\tilde{a} &= -\lambda^2\, r_2^2, \\
	\tilde{b} &= -\lambda\, \kappa_1\, r_2 - \lambda^2\, r_1\, r_2, \\
	\tilde{c} &= \kappa_0\, (\kappa_2 - \lambda\, r_2) - \tfrac{1}{4}(\kappa_1 + \lambda\, r_1)^2.
\end{align}
Figure~\ref{fig:Preqcrit_vs_lambda} visualizes how $P_\mathrm{crit}$ depends on $\lambda$.

\subsection{Classical Soft Constraints} \label{sec:appendixLambdaEb}
A classical reformulation of (\ref{eq:COP_cost}), (\ref{eq:OCP_terminal}), and (\ref{eq:OCP_state_bounds}) using soft constraints for the state bounds looks as follows:
\begin{equation}
	\begin{aligned}
	    \underset{\tilde{\mathbf{u}},\,\epsilon_1,\,\epsilon_2,\,\epsilon_3}{\text{min}} \int_{t_0}^{t_0+H_p} (P_f + a_1\cdot & \epsilon_1 + a_2\cdot \epsilon_2) \, dt + a_3\cdot \epsilon_3\\
		E_{b_\mathrm{min}}-\epsilon_1 & \leq E_b(t) \leq E_{b_\mathrm{max}}+\epsilon_2\\
		E_b(t_0+H_p) & \geq E_b^\mathrm{RTG}-\epsilon_3\\
		\epsilon_1,\epsilon_2,\epsilon_3  &\geq 0,
	\end{aligned}
\end{equation}
whereas its Hamiltonian function reads:
\begin{equation}
	\begin{aligned}
	    H(E_b, \tilde{\mathbf{u}}, \lambda) & = P_f -\lambda\cdot P_{s_b}\\
	    &+ a_1 \cdot (E_{b_\mathrm{min}}-E_b) +  a_2 \cdot (E_b - E_{b_\mathrm{max}})\\
	    & \quad \quad \quad \quad \quad \quad \quad \quad \,\, + a_3 \cdot (E_b^\mathrm{RTG}-E_b).
	\end{aligned}
\end{equation}
Note that the time-dependencies for $a_1, a_2, \epsilon_1, \epsilon_2, \lambda$ are omitted above. Using PMP, the necessary conditions for optimality regarding the costate are:
\begin{equation}
	\begin{aligned}
	    \dot{\lambda}(t) & = -\nabla_{E_b} H(E_b^\star, \tilde{\mathbf{u}}^\star, \lambda(t))\\
	    \lambda(T) & = \nabla_{E_b} \Psi(E_b^\star(T)).
	\end{aligned}
\end{equation}
If slacking is required to obtain a feasible problem formulation, the soft constraints's influence on the costate is one (or a combination) of the following:
\begin{equation}
	\begin{aligned}
	\dot{\lambda}(t) & = 
		\begin{cases}
		a_1(t) \hspace{1.4cm} \text{if $\epsilon_1(t) > 0$}\\
		-a_2(t) \hspace{1.1cm} \text{if $\epsilon_2(t) > 0$}
		\end{cases}\\
	\lambda(T) & = a_3 \hspace{2.05cm} \text{if $\epsilon_3 > 0$}.
	\end{aligned}
\end{equation}

\bibliographystyle{unsrt}
\begingroup
\raggedright
\bibliography{Learning_MPC_bib,JabRefBibTexDatabase}           %
\endgroup

\end{document}